\documentclass{article}
\usepackage[utf8]{inputenc}
\usepackage{authblk}
\usepackage{setspace}
\usepackage[margin=1.25in]{geometry}
\usepackage{graphicx}
\graphicspath{ {./figures/} }
\usepackage{subcaption}
\usepackage{amsmath}
\usepackage{cite}

\title{High-flux 100-kHz attosecond pulse source driven by a high average power annular laser beam}

\author[1,*,$\dag$, $\ddag$]{Peng Ye}
\author[1,2,$\dag$]{L\'{e}n\'{a}rd Guly\'{a}s Oldal}
\author[1]{Tam\'{a}s Csizmadia}
\author[1]{Zolt\'{a}n Filus}
\author[1]{T\'{i}mea Gr\'{o}sz}
\author[1]{P\'{e}ter J\'{o}j\'{a}rt}
\author[1]{Imre Seres}
\author[1]{Zsolt Bengery}
\author[1]{Barnab\'{a}s Gilicze}
\author[1,2]{Subhendu Kahaly}
\author[1,3]{Katalin Varj\'{u}}
\author[1,3,**]{Bal\'{a}zs Major}

\affil[1]{ELI-ALPS, ELI-HU Non-Profit Ltd., Wolfgang Sandner utca 3., Szeged, H-6728, Hungary.}
\affil[2]{Institute of Physics, University of Szeged, D\'{o}m t\'{e}r 9, Szeged 6720, Hungary.}
\affil[3]{Department of Optics and Quantum Electronics, University of Szeged, D\'{o}m t\'{e}r 9, Szeged 6720, Hungary.}
\affil[*]{Corresponding author. peng.ye@eli-alps.hu}
\affil[**]{Corresponding author. balazs.major@eli-alps.hu}
\affil[$\dag$]{These authors contributed equally to this work.}
\affil[$\ddag$]{Present address: LIDYL, CEA, CNRS, Universit\'{e} Paris-Saclay, CEA Paris-Saclay, Gif sur Yvette, France.}

\date{}

\begin{document}
	
	\maketitle
	
	\begin{abstract}
		High-repetition-rate attosecond pulse sources are indispensable tools of time-resolved studies of electron dynamics, such as coincidence spectroscopy and experiments with high demands on statistics or signal-to-noise ratio, especially in case of solid and big molecule samples in chemistry and biology. Although with the high-repetition-rate lasers such attosecond pulses in a pump-probe configuration are possible to achieve, until now only a few such light sources have been demonstrated. Here, by shaping the driving laser to an annular beam, a 100-kHz attosecond pulse train (APT) is reported with the highest energy so far (51 pJ/shot) on target (269 pJ at generation) among the high-repetition-rate systems ($>$ 10 kHz) in which the attosecond pulses were temporally characterized. The on-target pulse energy is maximized by reducing the losses from the reflections and filtering of the high harmonics, and an unprecedented $19\%$ transmission rate from the generation point to the target position is achieved. At the same time, the probe beam is also annular, and low loss of this beam is reached by using another holey mirror to combine with the APT. The advantages of using an annular beam to generate attosecond pulses with a high average power laser is demonstrated experimentally and theoretically. The effect of nonlinear propagation in the generation medium on the annular-beam generation concept is also analyzed in detail.
	\end{abstract}
	
	\section{Introduction}
\begin{figure}
	\centering
	\includegraphics[scale=0.7]{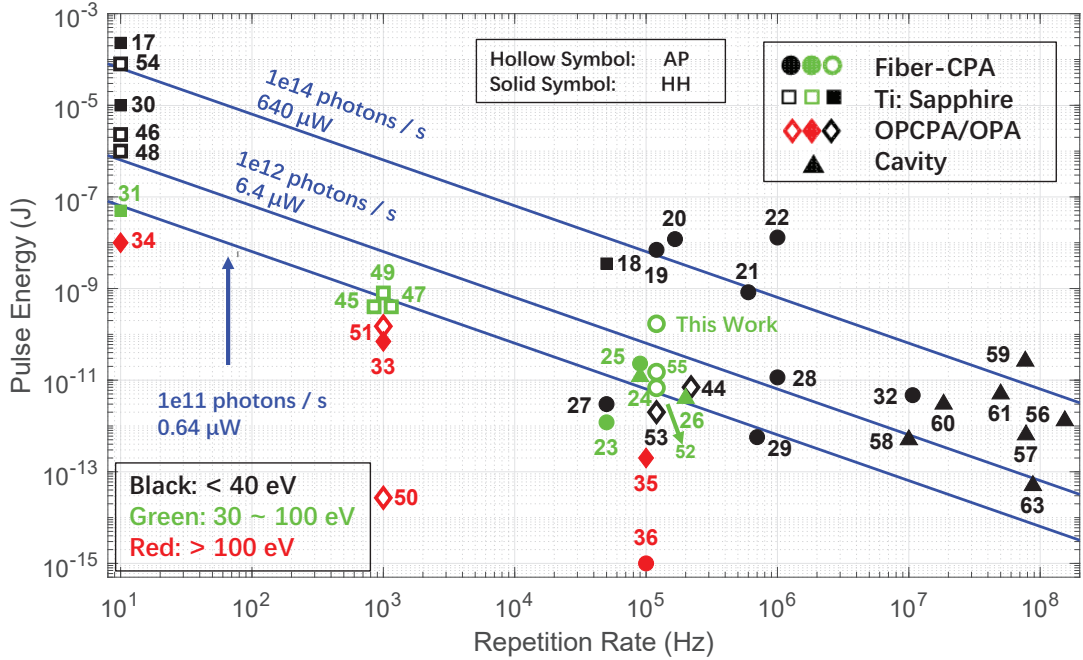}%
	\caption{\label{ATTOSOURCE} Typical energies per shot of the high-order harmonics at the repetition rates ranging from 10 Hz to 200 MHz. Hollow symbols show the attosecond pulses (AP) with measured attosecond duration. Solid symbols show high harmonics (HH) without temporal characterization. Circle/Square/Diamond/Triangle represent the technologies used: Fiber-CPA/Ti:Sapphire/OPA/Cavity. The three blue lines show the photon numbers at 40 eV at different repetition rates. The black/green/red colors differentiate by the photon energy range covered.}
\end{figure}

Since the first experimental realizations of an attosecond pulse train (APT) \cite{Paul01062001} and an isolated attosecond pulse (IAP) \cite{hentschel2001attosecond} at 1 kHz in 2001, attosecond pulses have been widely used to investigate electron dynamics in gases \cite{itatani2004tomographic}, liquids \cite{luu2018extreme} and solids \cite{ghimire2019high}. Nowadays, more effort is put towards scaling up the flux of APTs and IAPs using the lasers of high repetition rate. One route is the multi-pass high-harmonic generation (HHG) in a laser cavity or a resonant enhancement cavity \cite{pupeza2021extreme}, in which cases the lasers with a low pulse energy (nJ $\sim$ $\mu$J) and a very high repetition rate ($>$ MHz) are used. Another route is the single-pass HHG \cite{hadrich2016single}. In this case, since the laser energy is close to that of a typical kHz system ($\sim$ mJ), one can keep the same pump-probe ability of the kHz system, and at the same time increase the repetition rate to 100 kHz.  For the applications in which it is crucial to avoid space charge effects, such as the photoemission spectroscopy \cite{zheng2021ultrafast}, and in time-resolved coincidence measurements which require few events in each laser shot \cite{frasinski2013dynamics, cattaneo2018attosecond}, in order to achieve a high signal-to-noise ratio, a high repetition rate and a moderate attosecond pulse energy are preferred. Furthermore, because the time necessary for data collection can be shortened, high repetition rate is beneficial in a wide range of experiments such as coherent diffraction imaging \cite{miao2015beyond}, transient absorption \cite{johnson2016measurement}, and attosecond pump-probe spectroscopy \cite{ramasesha2016real}. For example, it will enhance the scope of single particle structural dynamics studies \cite{gorkhover2016femtosecond} and allows to investigate the newly emerged Schr\"{o}dinger cat states using strong laser fields \cite{rivera2021new, lewenstein2021generation}. Thanks to the continuous development of laser technology, high-repetition-rate and high-average-power lasers have become available, and as a result there is a continuous increase in the achievable photon flux. In this work, we call the high-order harmonics with measured attosecond temporal duration as attosecond pulses, and call those without temporal characterization as high harmonics. We make this distinction because temporal characterization is a demonstration of attosecond pump-probe capability. As shown in Figure \ref{ATTOSOURCE}, the single-pass HHG can provide the high harmonics \cite{nayak2018multiple, wang2015bright, klas2016table, comby2019cascaded, hadrich2014high, klas2021ultra, hadrich2011generation, lorek2014high, rothhardt2016high, rothhardt2016high, harth2017compact, klas2018annular, keunecke2020time, chiang2015boosting, takahashi2002generation, takahashi2004low, hadrich2015exploring, johnson2018high, fu2020high, chevreuil2021water, gebhardt2021bright} up to tens of nJ per shot at 1 MHz by using powerful driving lasers with an average power up to $\sim$ 100 W \cite{mero201843, nagy2019generation, storz2017parametric, zouflat, natile2019cep, young2018roadmap, toth2020sylos}, and the attosecond pulses \cite{mikaelsson2020high, goulielmakis2008single, takahashi2013attosecond, fabris2015synchronized, manschwetus2016two, timmers2016polarization, cousin2017attosecond, li2019double, ye2020attosecond, osolodkov2020generation, makos2020alpha, witting2021generation} up to hundreds of pJ per shot at 100 kHz. Intra-cavity HHG can deliver the high harmonics with the repetition rates up to hundreds of MHz \cite{cingoz2012direct, pupeza2013compact, ozawa2015high, porat2018phase, saule2019high, lee2011optimizing, carstens2016high, corder2018ultrafast}.

For single-pass HHG, two difficulties emerge when the high-harmonic generation (HHG) flux is scaled up by increasing the average power of such high-repetition-rate driving lasers. In a typical attosecond beamline, the incident laser beam is divided into a driving beam for HHG and a probe beam used in extreme ultraviolet - infrared (XUV-IR) pump-probe schemes for either the temporal characterization of attosecond pulses or for studying dynamics in the attosecond regime. The first challenge is to remove the high-average-power residual laser after generation process without attenuating the attosecond pulses drastically. Conventionally, a metal foil with a thickness of a few hundred nanometers is used to block the residual driving laser, allowing the transmission of the attosecond pulses with some losses. This method fails when the laser power increases because the thin foil is destroyed. The second difficulty arises due to the probe beam, the energy of which should be high enough for probing the system, such as what is needed for temporal characterization of the attosecond pulses. Most energy of the laser is given to the driving beam and only a small portion is in the probe beam. Conventionally, a holey mirror is used to combine the high-order harmonics which transmits through the central hole and the probe beam which is reflected. Because of a prominent energy loss due to the central hole, the energy of the probe beam available at target is even lower.     

In this work, by shaping the driving laser to an annular beam, we present a record-high APT energy in our 100-kHz attosecond beamline and demonstrate the advantages of utilizing annular beams to generate and characterize attosecond pulses with high average power IR laser beams. We show a proper technique with which the residual annular driving beam can be easily filtered out after HHG. We use an IR probe beam which is also annular at the holey recombination mirror. In this way it can be combined with the attosecond pulses with low loss, so an even bigger fraction of the driving laser energy can be used for HHG, which altogether results in a higher XUV flux. With this configuration we demonstrate 51 pJ energy of the 100 kHz attosecond pulses at the experimental target position. As shown in Figure \ref{ATTOSOURCE}, to the best of our knowledge this is the highest energy of attosecond pulses with the temporal characterization at the target achieved with high-repetition-rate systems. The $19.0\%$ transmission rate from generation to target is also the highest rate achieved so far in cases of using a high average power laser in the 100 W regime. The full width at half-maximum (FWHM) duration of the APTs was measured to be 166 as. 
	
\section{Methods}
We used a 100-kHz fiber laser system as the input to this beamline to drive the HHG process (see Section 1 in the supplementary material for the details). 

\begin{figure*}
	\centering\includegraphics[scale=0.45]{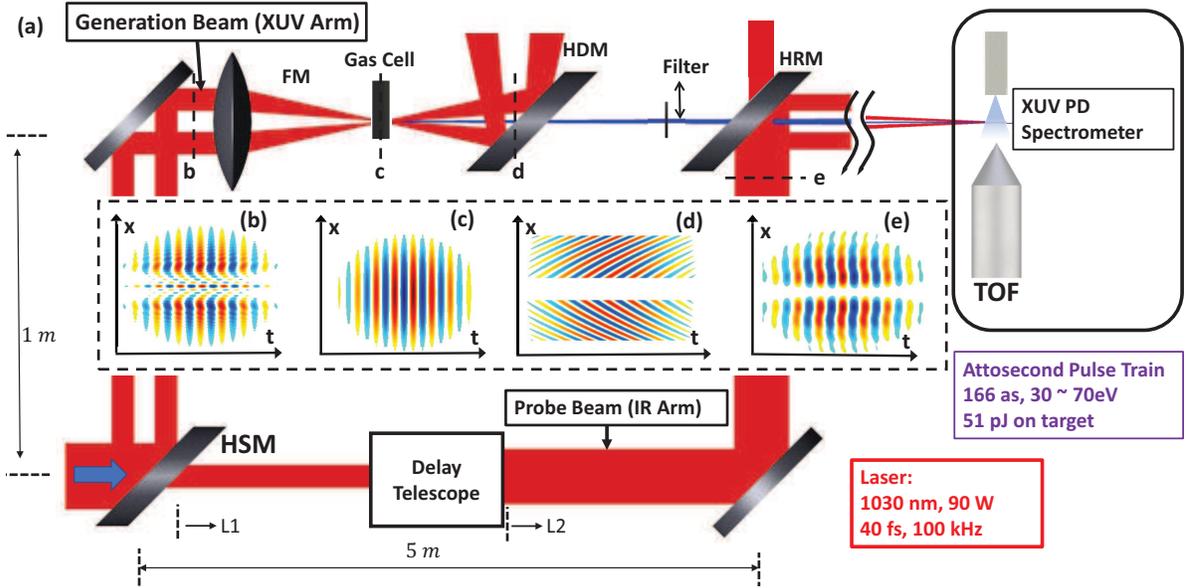}%
	\caption{\label{BeamPathRayTracing} Schematic representation of attosecond pulse generation and measurement using an annular beam. (a) Ray tracing of the beam path. HSM: holey splitting mirror. FM: focusing mirror. HDM: holey dump mirror. HRM: holey recombination mirror. The red beam is the fundamental laser beam. The blue beam is the high-order harmonic beam. Figures (b) to (d) show the simulated electric field of the generating pulsed laser beam at different positions. (b) in the front of the FM, (c) at focus, (d) in the front of the HDM. (e) is the electric field of the probe beam at the HRM. TOF: time-of-flight electron spectrometer. PD: photodiode. The hole sizes of the HSM, HDM and HRM are the same (6 mm in diameter).}
\end{figure*}

Figure \ref{BeamPathRayTracing} is a schematic of the beamline analyzed in this work which illustrates the XUV-IR pump-probe configuration. Figure \ref{BeamPathRayTracing}(a) shows the results of ray tracing of the beam path based on geometric optics. Figures \ref{BeamPathRayTracing}(b) to (e) show the electric field calculated using the paraxial wave equation at different positions. Red and blue colors represent positive and negative amplitudes respectively, while the white color show zero amplitude. The laser beam is magnified to 11 mm FWHM before reaching the beamline (blue arrow in Figure \ref{BeamPathRayTracing}). A holey splitting mirror (HSM) splits the input laser into a reflected annular beam (generation beam) and a transmitted central beam (probe beam). The generation beam is focused by the focusing mirror (FM) with the focusing length of 0.9 m onto a spot at the gas cell to generate high-order harmonics, as shown in Figure \ref{BeamPathRayTracing}(c). The gas cell is a home-made water-cooled gas cavity designed to be used in combination with high average power laser beams \cite{FilusZoltan2022}. This generation beam propagates to an annular shape with a hole in the center after the HHG shown in both Figure \ref{BeamPathRayTracing}(a) and (d), so it can be reflected off fully by another holey mirror, the holey dump mirror (HDM). The XUV goes through the center without any attenuation. It must be noted that based on ray tracing the generation beam is perfectly annular everywhere except at the focus in Figure \ref{BeamPathRayTracing}(a). However, wave propagation gives a different behavior. The shape of the generation beam is a diffraction pattern evolving along the beam path with substantial energy in the center. A typical pattern is shown in Figure \ref{BeamPathRayTracing}(b). The perfect annular shape with no energy in the center can be observed only in a small range, which is the suitable place for the HDM. The probe beam goes through the central hole of the HSM. After the delay stage and the telescope, the magnified and delayed IR beam is combined with the high-order harmonics using a holey recombination mirror (HRM). As indicated in Figure \ref{BeamPathRayTracing}(a), there is some energy loss in the probe beam after the HRM as the transmitted central part is lost through the hole. However, wave propagation predicts an annular shape of the probe beam at the HRM, shown in Figure \ref{BeamPathRayTracing}(e), therefore the energy loss can be avoided. After the recombination, the high-order harmonics and the probe beam are focused into the time-of-flight (TOF) electron spectrometer for the temporal characterization of the XUV. An XUV photodiode (PD) and an XUV spectrometer placed after the TOF are used to measure the energy and the spectrum of the XUV. The XUV beam path and the measurement of the flux can be found in the supplementary material (Figure S1 and S2 in section 2).

The propagation of the infrared beam is simulated using the Huygens-Fresnel integral and paraxial wave equation without the source term in free space \cite{siegman1986lasers, ye2014minimizing}(See section 3 for the details in Supplementary Material). While analyzing the spatial profile of the beams, we found that in the studied aspects a monochromatic beam and a pulse yield the same conclusion, so in the following we will only consider the monochromatic beam for simplicity. In the following, we will describe the evolution of the generation beam and the probe beam based on the wave equation, and show the proper arrangement of optics for dumping the generation beam and for recombining the XUV and probe beams.

\section{Results}
\subsection{Generation Beam (XUV Arm)}

In order to block the high-average-power residual generation beam, two methods have been proposed so far for high average power driven HHG: (i) one method is to use plates with special coating to reflect the attosecond pulses and transmit the driving laser. The reflection of two fused silica plates is as low as $17\%$ at 30 eV, and the attosecond pulses retain only $10\%$ of their energy after filtering \cite{hadrich2014high}. Furthermore, the coating must be individually designed to fit the laser spectrum, and it is challenging especially in the case of few-cycle lasers with a broad spectrum. The other method (ii) is to use an annular beam to generate high-order harmonics \cite{peatross1994high, kuhn2017eli}. The annular beam converges at focus to generate high harmonics, and becomes annular again, so it can be reflected off easily by a holey mirror or blocked by a holey plate after HHG. Attosecond pulses have already been generated and characterized using this approach with a 1 kHz laser \cite{mairesse2003attosecond} having a much lower average power (1 W). Generally, a small portion of the driving laser always co-propagates with the XUV, so the residual driving laser beam cannot be fully blocked. In case a low power laser is used, this small portion can be neglected. However, with the increase of laser power, this portion will become stronger, and it must be considered. In the high-repetition-rate regime of $\sim 100$ kHz, several laboratories have used annular laser beams to generate high-order harmonics \cite{klas2018annular, gaumnitz2018extreme}, while the measurement of the attosecond temporal duration was only reported in our previous work at ELI-ALPS \cite{ye2020attosecond}.

\begin{figure}
	\centering\includegraphics[scale=0.5]{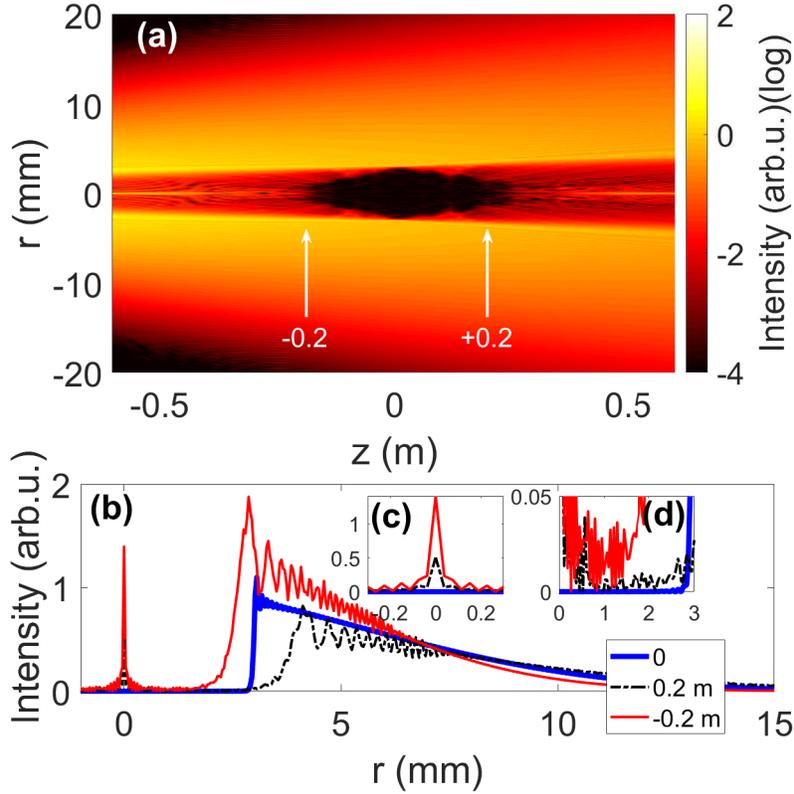}%
	\caption{\label{BeamPropagationHDM} (Simulation) (a) Beam propagation after the gas cell using paraxial wave equation. z = 0 is set as the image plane of the FM with the HSM as the object. (b) Radial intensity profile of the beam at different z positions. Blue: z = 0, the image plane of the HSM. Black: z = 0.2 m. Red: z = -0.2 m. The positions of z = $\pm$ 0.2 m are marked with the white arrows in (a). (c) The central part of the beam ( around r = 0 ). (d) The intensity distribution inside the hole.} 
\end{figure}

\begin{figure}
	\centering\includegraphics[scale=0.4]{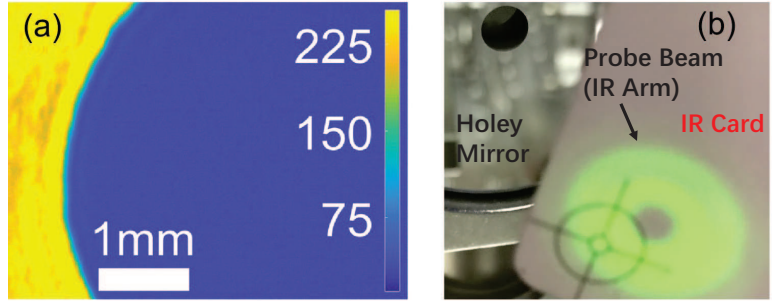}%
	\caption{\label{BeamEx} (Experiment) (a) The measured beam profile of the residual generation beam at the HDM using a CMOS camera. (b) The beam profile recorded on an IR card of the probe beam in front of the HRM.  }
\end{figure}

Figure \ref{BeamPropagationHDM} shows beam propagation after the gas cell for a monochromatic beam of 1030-nm wavelength. In Figure \ref{BeamPropagationHDM}(a) we observe that at the positions of $z < -0.2$ m the beam profile considerably differs from the geometrically expected central shadow, exhibiting substantial intensity of diffraction rings. In the range from z = -2 m to z = 0, the beam evolves gradually from the diffraction pattern to an ideal annular beam as predicted by geometrical optics. At the position of z = 0, i.e., the image plane of the HSM (the FM mirror serving as the imaging optic), the beam has a perfect annular shape with no light in the center, as can be seen in Figure \ref{BeamPropagationHDM}(c) and (d). After further propagation towards  $z > 0$, the beam again shows substantial diffraction in its profile. In Figure \ref{BeamPropagationHDM}(a) and (b) we observe the Arago spot before and after the image plane of the HSM ( z = 0 ). In addition, at the position of z = 0 there is a circular on-axis area of the beam with no light inside it, as shown in Figure \ref{BeamPropagationHDM}(c) and (d). However, when z = 0.2 m and z = –0.2 m, a considerable amount of light can still be observed in the center. When low average power laser beams are used \cite{mairesse2003attosecond}, the Arago spot and the diffraction rings do not have sufficient intensity to cause practical issues. However, when the average power of the laser is increased, these diffraction rings must be considered, since they can damage the optical elements and detectors and can produce unwanted noise in the signal. In order to reflect such a beam fully with a holey mirror, the mirror must be put at the image plane. Practically, the mirror must be placed in the diamond-shaped dark area in Figure \ref{BeamPropagationHDM}(a). In our case, the HDM can be located within 10 cm around z = 0. We have also simulated the cases when focusing mirrors with different focal lengths between 0.5 m and 3 m are used. The results have shown that the optimal position range is not directly related to the focal length, and it is between 10 cm to 30 cm in the case of all studied focal lengths. Researchers aiming to design such a beamline must simulate beam propagation using the wave equation instead of ray tracing to find the appropriate range, and must experimentally measure the beam profile to check the correctness of the positions. In order to directly record the beam profile, we put a CMOS sensor at the position of the HDM, and measured the beam profile at low power (1 W) and atmospheric pressure. As shown in Figure \ref{BeamEx}(a), the beam is perfectly annular, so it can be fully reflected by the HDM.

\begin{figure}
	\centering\includegraphics[scale=0.7]{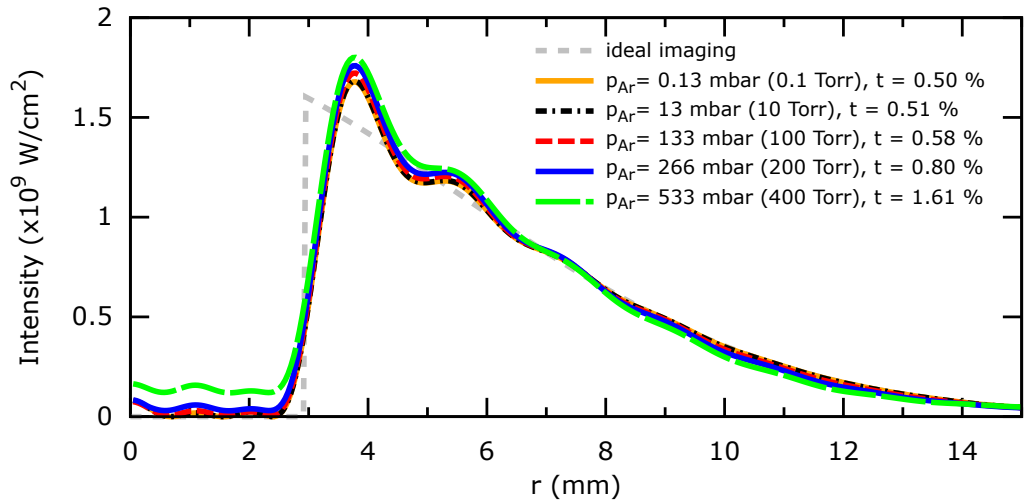}%
	\caption{\label{Ionization_HDM} (Simulation) The beam profile of the driving laser at the HDM (see position in Figure \ref{BeamPathRayTracing}) at different pressures of Ar medium ($p_{Ar}$) in the gas cell. The ideal imaging (dashed gray curve) shows the case of free propagation in vacuum without diffraction on any obstacle. The transmission percentages (t) in the legend provide the ratio of the beam energy transmitted through the central hole of the HDM with respect to the input beam energy reaching the beamline. The solid blue curve corresponds to the experimental case.}
\end{figure}

In the above discussion, we only considered the pulse propagation in vacuum. If the spatio-temporal distribution of the generation beam is not prominently changed by the medium, i.e., from all the possible nonlinear effects only the process of HHG takes place, the conclusion of this work is valid without restrictions. In real experiments, if HHG works under the usual phase matching conditions, where the ionization is lower than the critical ionization rate (usually less than few percent) \cite{chang2016fundamentals}, the driving laser can be considered unmodified by the gas, and our conclusions are not affected by these effects. However, in the case of high ionization, the shape of the laser beam during propagation will be modified by the electrons in the medium  \cite{johnson2018high, rivas2018propagation, major2019effect, major2020propagation}, and the far-field shape of the laser beam profile is expected to change relevantly. To analyze the effect of ionization of the generation medium, we carried out simulations. To analyze the effect of ionization of the generation medium, we carried out additional simulations (See section 4 in Supplementary Material for the details). In the simulations, a gas cell with 4-mm length and 1.2-mm diameter was put at the position of the laser focus, matching the experimental conditions. We changed the pressure of argon and calculated the beam profile at the position of the HDM in Figure \ref{BeamPathRayTracing}. As shown in Figure \ref{Ionization_HDM}, by increasing the pressure, the transmitted energy also increases. Using the same parameters as in the experiments, when the pressure is $\sim$ 200 mbar in the gas cell, the transmitted energy is below $1\%$.  In the case of using 100-W laser,  the transmitted power is below 1 W (the same level of 1-kHz system) and can be safely blocked by a metallic filter. At higher pressure of $\sim$ 500 mbar and higher free-electron density, the transmitted energy is still below $ 2\%$. It should be noted that there is a certain percentage of beam energy always transmitted through the hole of the HDM because the cell aperture acts as a spatial filter distorting ideal imaging conditions (see details in section 4 for the Supplementary material). Also, the almost unchanged transmission percentages ($t$) and beam profiles in Figure \ref{BeamPropagationHDM} up to a medium pressure of $p_{Ar}\sim$ 10 mbar suggests that at these pressures with our focused laser intensities the situation is identical to propagating in vacuum. In a recent theoretical work, Cheng Jin et al. also investigated HHG in the overdriven regime (high ionization) using an annular beam and indicated that XUV and IR can separate in the far field \cite{jin2020optimal}.

\subsection{Probe Beam (IR Arm)}

\begin{figure}
	\centering\includegraphics[scale=0.5]{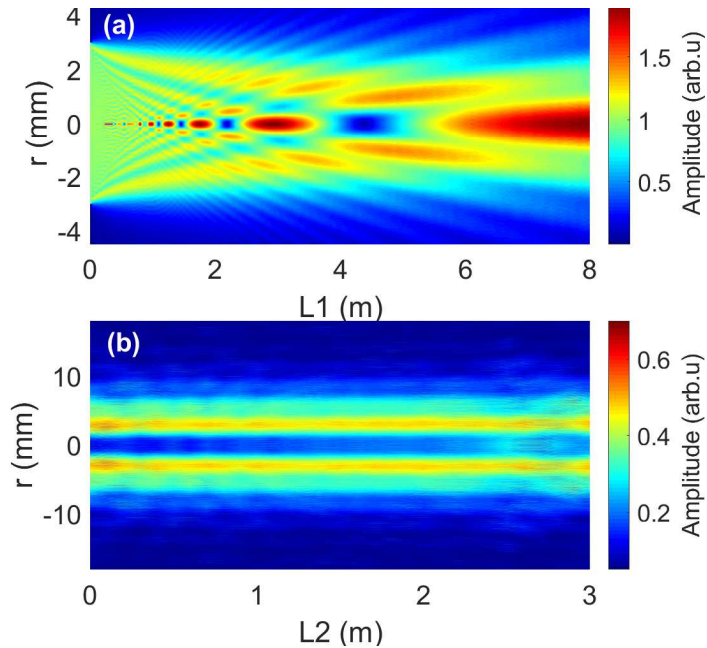}%
	\caption{\label{BeamHDM} (Simulation) (a) The amplitude of the probe beam after the HSM with a 6 mm diameter hole. L1 = 0 is the position of the HSM shown in Figure \ref{BeamPathRayTracing}. The telescope is positioned to image an annular part of the propagating beam. The entrance of the telescope is around the position of L1 = 4.1 m. (b) The amplitude of the annular probe beam after the telescope. L2 = 0 is the exit of the telescope, also shown in Figure \ref{BeamPathRayTracing}. The amplitudes in the two pictures are in linear scale.}
\end{figure}

To match the focus of the attosecond and IR pulses in the pump-probe setup, the probe beam is recombined with the attosecond pulses by a holey mirror. The attosecond pulses propagate through the central hole, while the IR is reflected. This scheme wastes the central part of the IR probe causing a relevant loss in its energy during recombination. Although the average power of the laser is high, the energy of the individual pulses is low (below mJ or even 100 $\mu$J in most of the currently available systems). In order to perform a reconstruction of attosecond beating by interference of two-photon transitions (RABBITT) \cite{Paul01062001} or a streaking measurement \cite{hentschel2001attosecond}, the laser intensity of the probe pulse must be above $10^{11}$ Wcm$^{-2}$, so these losses need to be minimized.

In our beamline, shown in Figure \ref{BeamPathRayTracing}(a), the probe beam is magnified by a telescope and then combined with the high-order harmonics using the HRM, where a substantial amount of energy in the center would be lost. However, diffraction allows for system optimization. As shown in Figure \ref{BeamPathRayTracing}(a), the transmitted probe beam from the HSM (L1 = 0) evolves as a diffraction pattern. The central intensity exhibits an oscillating behavior along the laser propagation direction. For certain positions of the HRM, e.g., at z = 3 m, most energy would be lost through the hole. However, by positioning it at z = 4.5 m, almost the entire energy of the beam could be preserved after reflection, since the transmitted central part is a hole with a low portion of energy. However, as the position of HRM cannot be set completely arbitrarily in most beamlines, therefore we use a telescope to position the annular profile of the probe beam to a suitable geometrical position.

We build the telescope at the position of L1 = 4.1 m in Figure \ref{BeamHDM}(a), where the probe beam exhibits an annular shape. The telescope has threefold magnification, and the propagation distance is virtually reduced by 70 cm as a result of imaging. L2 = 0 is defined as the output of the telescope in Figure \ref{BeamHDM}(b) (see also Figure \ref{BeamPathRayTracing}). The magnified beam propagates further and keeps its annular shape within 2 meters, as shown in Figure \ref{BeamHDM}(b). Figure \ref{BeamEx}(b) shows the annular beam profile of the probe beam on an IR card in front of a holey mirror. In our experiment the hole diameter of HRM was 6  mm, and the loss due to reflection was $15\%$. According to our simulations, this loss can be decreased to $3.5\%$ by reducing the hole diameter to 4 mm.

\subsection{Attosecond Pulse Duration Measurement}
\begin{figure}
	\centering\includegraphics[scale=0.5]{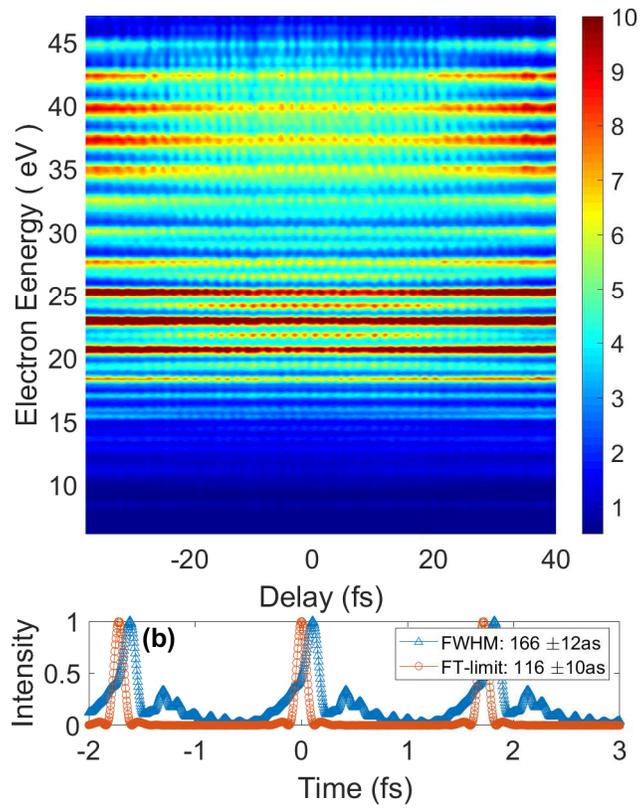}%
	\caption{\label{R_Trace} (Experiment) (a) Measured RABBITT trace (linear scale). (b) The FWHMs of the fourier-transform-limited pulse(red) and the reconstructed pulse (blue). Both the red line and blue line were normalized to the peak of the red line. }
\end{figure}

To demonstrate the performance of our system optimized according to the description above, high-order harmonics were generated in a 4-mm gas cell filled with 200-mbar argon gas. The generated harmonic beam propagates through a 100-nm aluminum (Al) foil, and combines with the delayed IR beam. The two beams are focused to ionize neon (Ne) gas from a gas jet in front of a TOF spectrometer that collects the emitted electrons. By changing the delay between the two beams we can record the delay dependent electron kinetic energy spectrogram, i.e., RABBITT trace, shown in Fig. \ref{R_Trace}(a). The photon energy covered by the APT was between 30 eV and 70 eV. As a result, the electron kinetic energies were ranging from 8 eV to 48 eV, obtained by subtracting the 21.56-eV ionization potential of Ne. The whole temporal range of the trace is approximately $\sim$ 70 fs, which is consistent with the 40-fs duration (FWHM) of the independent measurement of the driving laser \cite{ye2020attosecond}. The reconstruction gave an average FWHM duration of 166$\pm$12 as of the attosecond pulses in the APT, as shown in Fig. \ref{R_Trace}(b) (blue line). After the TOF, a photodiode was inserted in the beam path to measure the energy of high-order harmonics at the target position. The pulse energy was measured to be 51.0$\pm$3.1  pJ. The pulse energy at generation was calculated to be 269.0 pJ. Further details of the beamline and the laser system can be found in our previous works \cite{kuhn2017eli, ye2020attosecond}. Details of the energy measurement of the high-order can be found in the Figure S1 and Figure S2 in Section 2 of the supplementary material. Compared to our previous work \cite{ye2020attosecond}, we optimized the experimental conditions by using a new water-cooled gas cell, an extensive parametric optimization of phase matching conditions and improving the stability of the laser. We reached a five-fold decrease in necessary integration time while also improving the signal-to-noise ratio. These improved results demonstrate the possibility to carry out attosecond pump-probe measurements at 100 kHz repetition rate with our beamline at XUV fluxes not available before.
 
\section{Conclusion}
In conclusion, in this work we have reported the generation and temporal characterization of attosecond pulses using the high average power HR laser of ELI-ALPS. In our approach, a holey mirror is used to split the laser into two independent beams. The reflected annular beam is used for attosecond pulse generation, while the transmitted central part serves as the probe beam for experiments and temporal characterization. After HHG, the generation beam becomes annular again upon further propagation. As predicted by wave optics, if a holey mirror is placed at a proper position, the residual annular IR beam can be almost fully reflected, and the harmonics can transmit through the central hole. This way the possible damage of the optics and detectors by the residual generating IR beam can be avoided, and the unwanted background in the signal can also be suppressed. Since the diffracted central probe beam also becomes annular, it can be recombined with the XUV beam via reflection on a holey mirror with minimal energy loss, provided that this mirror is placed at the correct position. This ensures a sufficiently intense probe beam for pump-probe experiments such as RABBITT or streaking measurements. These assumptions have been verified both by experiments and simulations, and have provided guidance in finding the proper positions of the key mirrors in our beamline. As a result, we could optimize HHG and delivered 51.0 pJ attosecond pulse trains with an average duration of 166 as to the target position after transmission through a 100-nm thick Al metal foil. This is the highest pulse energy of attosecond pulses with temporal characterization achieved so far on target using a laser with a repetition rate higher than 10 kHz and an average power in 100 W regime. As a future step, we plan to compress the laser pulses further to a few-cycle duration \cite{shestaev2020high}. We expect that this approach will even increase the conversion efficiency and the flux of the attosecond pulses. With this 100-kHz high-energy attosecond pulse, we believe that many experiments which need both the high repetition rate and enough energy can be performed now, especially for the studies of solid samples and big molecules.

\section*{Acknowledgments}
We thank the groups of Mauro Nisoli and Luca Poletto for the help provided in developing the beamline and the discussions. We also thank Harshitha Nandiga Gopalakrishna, Miklos F\"{u}le, and Amelle Za\"{i}r for the early contributions to the implementation of this beamline. We are grateful to Valer Tosa for the availability of the macroscopic high-harmonic generation simulation code. We acknowledge KIF\"{U} for awarding us high-performance computing access to resource based in Hungary.

\subsection*{Author Contributions} 
P.Y., with the input from B.M., conceived the idea. K.V. initiated the project of the beamline. B.M., S.K., and K.V. supervised the project. P.Y, L.G.O., T.C., Z.F., T.G., and B.M. operated the beamline and performed the experiments. P.Y. and B.M. performed the simulations, and with L.G.O. analyzed the experiment results. P.J., I.S., Z.B., and B.G. operated the laser system. P.Y. wrote the manuscript with the inputs from B.M, S.K., and K.V. All people contributed to the preparation of the manuscript.         

\subsection*{Funding}
The ELI-ALPS project (GINOP-2.3.6-15-2015-00001) is supported by the European Union and co-financed by the European Regional Development Fund. 

\subsection*{Conflicts of Interest}
The authors declare no conflict of interest.

\subsection*{Data Availability}
All data discussed in the article will be made available upon request.
	
	\section*{Supplementary Materials}
	Section 1: The details of the laser system.
	Section 2: The beampath of the XUV path (Figure S-1) and the flux measruement (Figure S-2).
	Section 3: The simulation of beam propagation in free space.
	Section 4: The simulation of beam propagation in ionized gas and the aperture effect.
	

\bibliographystyle{ieeetr}

\bibliography{sample}

\begin{thebibliography}{10}

\bibitem{Paul01062001}
P.~M. Paul, E.~S. Toma, P.~Breger, G.~Mullot, F.~Auge, P.~Balcou, H.~G. Muller,
  and P.~Agostini, ``Observation of a train of attosecond pulses from high
  harmonic generation,'' {\em Science}, vol.~292, no.~5522, pp.~1689--1692,
  2001.

\bibitem{hentschel2001attosecond}
M.~Hentschel, R.~Kienberger, C.~Spielmann, G.~Reider, N.~Milosevic, T.~Brabec,
  P.~Corkum, U.~Heinzmann, M.~Drescher, and F.~Krausz, ``Attosecond
  metrology,'' {\em Nature (London)}, vol.~414, no.~6863, pp.~509--513, 2001.

\bibitem{itatani2004tomographic}
J.~Itatani, J.~Levesque, D.~Zeidler, H.~Niikura, H.~P{\'e}pin, J.-C. Kieffer,
  P.~B. Corkum, and D.~M. Villeneuve, ``Tomographic imaging of molecular
  orbitals,'' {\em Nature}, vol.~432, no.~7019, pp.~867--871, 2004.

\bibitem{luu2018extreme}
T.~T. Luu, Z.~Yin, A.~Jain, T.~Gaumnitz, Y.~Pertot, J.~Ma, and H.~J.
  W{\"o}rner, ``Extreme--ultraviolet high--harmonic generation in liquids,''
  {\em Nature Communications}, vol.~9, no.~1, pp.~1--10, 2018.

\bibitem{ghimire2019high}
S.~Ghimire and D.~A. Reis, ``High-harmonic generation from solids,'' {\em
  Nature Physics}, vol.~15, no.~1, pp.~10--16, 2019.

\bibitem{pupeza2021extreme}
I.~Pupeza, C.~Zhang, M.~H{\"o}gner, and J.~Ye, ``Extreme-ultraviolet frequency
  combs for precision metrology and attosecond science,'' {\em Nature
  Photonics}, vol.~15, no.~3, pp.~175--186, 2021.

\bibitem{hadrich2016single}
S.~H{\"a}drich, J.~Rothhardt, M.~Krebs, S.~Demmler, A.~Klenke,
  A.~T{\"u}nnermann, and J.~Limpert, ``Single-pass high harmonic generation at
  high repetition rate and photon flux,'' {\em Journal of Physics B: Atomic,
  Molecular and Optical Physics}, vol.~49, no.~17, p.~172002, 2016.

\bibitem{zheng2021ultrafast}
W.~Zheng, P.~Jiang, L.~Zhang, Y.~Wang, Q.~Sun, Y.~Liu, Q.~Gong, and C.~Wu,
  ``Ultrafast extreme ultraviolet photoemission electron microscope,'' {\em
  Review of Scientific Instruments}, vol.~92, no.~4, p.~043709, 2021.

\bibitem{frasinski2013dynamics}
L.~Frasinski, V.~Zhaunerchyk, M.~Mucke, R.~J. Squibb, M.~Siano, J.~H. Eland,
  P.~Linusson, P.~Vd~Meulen, P.~Sal{\'e}n, R.~Thomas, {\em et~al.}, ``Dynamics
  of hollow atom formation in intense x-ray pulses probed by partial covariance
  mapping,'' {\em Physical Review Letters}, vol.~111, no.~7, p.~073002, 2013.

\bibitem{cattaneo2018attosecond}
L.~Cattaneo, J.~Vos, R.~Y. Bello, A.~Palacios, S.~Heuser, L.~Pedrelli,
  M.~Lucchini, C.~Cirelli, F.~Mart{\'\i}n, and U.~Keller, ``Attosecond coupled
  electron and nuclear dynamics in dissociative ionization of {H}$_2$,'' {\em
  Nature Physics}, vol.~14, no.~7, pp.~733--738, 2018.

\bibitem{miao2015beyond}
J.~Miao, T.~Ishikawa, I.~K. Robinson, and M.~M. Murnane, ``Beyond
  crystallography: Diffractive imaging using coherent x-ray light sources,''
  {\em Science}, vol.~348, no.~6234, pp.~530--535, 2015.

\bibitem{johnson2016measurement}
A.~Johnson, L.~Miseikis, D.~Wood, D.~Austin, C.~Brahms, S.~Jarosch,
  C.~Str{\"u}ber, P.~Ye, and J.~Marangos, ``Measurement of sulfur {L}$_{2, 3}$
  and carbon k edge xanes in a polythiophene film using a high harmonic
  supercontinuum,'' {\em Structural Dynamics}, vol.~3, no.~6, p.~062603, 2016.

\bibitem{ramasesha2016real}
K.~Ramasesha, S.~R. Leone, and D.~M. Neumark, ``Real-time probing of electron
  dynamics using attosecond time-resolved spectroscopy,'' {\em Annual Review of
  Physical Chemistry}, vol.~67, pp.~41--63, 2016.

\bibitem{gorkhover2016femtosecond}
T.~Gorkhover, S.~Schorb, R.~Coffee, M.~Adolph, L.~Foucar, D.~Rupp, A.~Aquila,
  J.~D. Bozek, S.~W. Epp, B.~Erk, {\em et~al.}, ``Femtosecond and nanometre
  visualization of structural dynamics in superheated nanoparticles,'' {\em
  Nature Photonics}, vol.~10, no.~2, pp.~93--97, 2016.

\bibitem{rivera2021new}
J.~Rivera-Dean, P.~Stammer, E.~Pisanty, T.~Lamprou, P.~Tzallas, M.~Lewenstein,
  and M.~F. Ciappina, ``New schemes for creating large optical schr{\"o}dinger
  cat states using strong laser fields,'' {\em Journal of Computational
  Electronics}, pp.~1--13, 2021.

\bibitem{lewenstein2021generation}
M.~Lewenstein, M.~Ciappina, E.~Pisanty, J.~Rivera-Dean, P.~Stammer, T.~Lamprou,
  and P.~Tzallas, ``Generation of optical schr{\"o}dinger cat states in intense
  laser--matter interactions,'' {\em Nature Physics}, vol.~17, no.~10,
  pp.~1104--1108, 2021.

\bibitem{nayak2018multiple}
A.~Nayak, I.~Orfanos, I.~Makos, M.~Dumergue, S.~K{\"u}hn, E.~Skantzakis,
  B.~Bodi, K.~Varju, C.~Kalpouzos, H.~Banks, {\em et~al.}, ``Multiple
  ionization of argon via multi-xuv-photon absorption induced by 20-gw
  high-order harmonic laser pulses,'' {\em Physical Review A}, vol.~98, no.~2,
  p.~023426, 2018.

\bibitem{wang2015bright}
H.~Wang, Y.~Xu, S.~Ulonska, J.~S. Robinson, P.~Ranitovic, and R.~A. Kaindl,
  ``Bright high-repetition-rate source of narrowband extreme-ultraviolet
  harmonics beyond 22 ev,'' {\em Nature Communications}, vol.~6, no.~1,
  pp.~1--7, 2015.

\bibitem{klas2016table}
R.~Klas, S.~Demmler, M.~Tschernajew, S.~H{\"a}drich, Y.~Shamir,
  A.~T{\"u}nnermann, J.~Rothhardt, and J.~Limpert, ``Table-top milliwatt-class
  extreme ultraviolet high harmonic light source,'' {\em Optica}, vol.~3,
  no.~11, pp.~1167--1170, 2016.

\bibitem{comby2019cascaded}
A.~Comby, D.~Descamps, S.~Beauvarlet, A.~Gonzalez, F.~Guichard, S.~Petit,
  Y.~Zaouter, and Y.~Mairesse, ``Cascaded harmonic generation from a fiber
  laser: a milliwatt xuv source,'' {\em Optics Express}, vol.~27, no.~15,
  pp.~20383--20396, 2019.

\bibitem{hadrich2014high}
S.~H{\"a}drich, A.~Klenke, J.~Rothhardt, M.~Krebs, A.~Hoffmann, O.~Pronin,
  V.~Pervak, J.~Limpert, and A.~T{\"u}nnermann, ``High photon flux table-top
  coherent extreme-ultraviolet source,'' {\em Nature Photonics}, vol.~8,
  no.~10, pp.~779--783, 2014.

\bibitem{klas2021ultra}
R.~Klas, A.~Kirsche, M.~Gebhardt, J.~Buldt, H.~Stark, S.~H{\"a}drich,
  J.~Rothhardt, and J.~Limpert, ``Ultra-short-pulse high-average-power
  megahertz-repetition-rate coherent extreme-ultraviolet light source,'' {\em
  PhotoniX}, vol.~2, no.~1, pp.~1--8, 2021.

\bibitem{hadrich2011generation}
S.~H{\"a}drich, M.~Krebs, J.~Rothhardt, H.~Carstens, S.~Demmler, J.~Limpert,
  and A.~T{\"u}nnermann, ``Generation of $\mu$w level plateau harmonics at high
  repetition rate,'' {\em Optics Express}, vol.~19, no.~20, pp.~19374--19383,
  2011.

\bibitem{lorek2014high}
E.~Lorek, E.~W. Larsen, C.~M. Heyl, S.~Carlstr{\"o}m, D.~Pale{\v{c}}ek,
  D.~Zigmantas, and J.~Mauritsson, ``High-order harmonic generation using a
  high-repetition-rate turnkey laser,'' {\em Review of Scientific Instruments},
  vol.~85, no.~12, p.~123106, 2014.

\bibitem{rothhardt2016high}
J.~Rothhardt, S.~H{\"a}drich, Y.~Shamir, M.~Tschnernajew, R.~Klas, A.~Hoffmann,
  G.~K. Tadesse, A.~Klenke, T.~Gottschall, T.~Eidam, {\em et~al.},
  ``High-repetition-rate and high-photon-flux 70 e{V} high-harmonic source for
  coincidence ion imaging of gas-phase molecules,'' {\em Optics Express},
  vol.~24, no.~16, pp.~18133--18147, 2016.

\bibitem{harth2017compact}
A.~Harth, C.~Guo, Y.-C. Cheng, A.~Losquin, M.~Miranda, S.~Mikaelsson, C.~M.
  Heyl, O.~Prochnow, J.~Ahrens, U.~Morgner, {\em et~al.}, ``Compact 200 k{H}z
  hhg source driven by a few-cycle {OPCPA},'' {\em Journal of Optics}, vol.~20,
  no.~1, p.~014007, 2017.

\bibitem{klas2018annular}
R.~Klas, A.~Kirsche, M.~Tschernajew, J.~Rothhardt, and J.~Limpert, ``Annular
  beam driven high harmonic generation for high flux coherent xuv and soft
  x-ray radiation,'' {\em Optics Express}, vol.~26, no.~15, pp.~19318--19327,
  2018.

\bibitem{keunecke2020time}
M.~Keunecke, C.~M{\"o}ller, D.~Schmitt, H.~Nolte, G.~M. Jansen, M.~Reutzel,
  M.~Gutberlet, G.~Halasi, D.~Steil, S.~Steil, {\em et~al.}, ``Time-resolved
  momentum microscopy with a 1 {MHz} high-harmonic extreme ultraviolet
  beamline,'' {\em Review of Scientific Instruments}, vol.~91, no.~6,
  p.~063905, 2020.

\bibitem{chiang2015boosting}
C.-T. Chiang, M.~Huth, A.~Tr{\"u}tzschler, M.~Kiel, F.~O. Schumann,
  J.~Kirschner, and W.~Widdra, ``Boosting laboratory photoelectron spectroscopy
  by megahertz high-order harmonics,'' {\em New Journal of Physics}, vol.~17,
  no.~1, p.~013035, 2015.

\bibitem{takahashi2002generation}
E.~Takahashi, Y.~Nabekawa, and K.~Midorikawa, ``Generation of 10-$\mu${J}
  coherent extreme-ultraviolet light by use of high-order harmonics,'' {\em
  Optics Letters}, vol.~27, no.~21, pp.~1920--1922, 2002.

\bibitem{takahashi2004low}
E.~J. Takahashi, Y.~Nabekawa, and K.~Midorikawa, ``Low-divergence coherent soft
  x-ray source at 13 nm by high-order harmonics,'' {\em Applied Physics
  Letters}, vol.~84, no.~1, pp.~4--6, 2004.

\bibitem{hadrich2015exploring}
S.~H{\"a}drich, M.~Krebs, A.~Hoffmann, A.~Klenke, J.~Rothhardt, J.~Limpert, and
  A.~T{\"u}nnermann, ``Exploring new avenues in high repetition rate table-top
  coherent extreme ultraviolet sources,'' {\em Light: Science \& Applications},
  vol.~4, no.~8, pp.~e320--e320, 2015.

\bibitem{johnson2018high}
A.~S. Johnson, D.~R. Austin, D.~A. Wood, C.~Brahms, A.~Gregory, K.~B. Holzner,
  S.~Jarosch, E.~W. Larsen, S.~Parker, C.~S. Str{\"u}ber, {\em et~al.},
  ``High-flux soft x-ray harmonic generation from ionization-shaped few-cycle
  laser pulses,'' {\em Science Advances}, vol.~4, no.~5, p.~eaar3761, 2018.

\bibitem{fu2020high}
Y.~Fu, K.~Nishimura, R.~Shao, A.~Suda, K.~Midorikawa, P.~Lan, and E.~J.
  Takahashi, ``High efficiency ultrafast water-window harmonic generation for
  single-shot soft x-ray spectroscopy,'' {\em Communications Physics}, vol.~3,
  no.~1, pp.~1--10, 2020.

\bibitem{chevreuil2021water}
P.-A. Chevreuil, F.~Brunner, S.~Hrisafov, J.~Pupeikis, C.~R. Phillips,
  U.~Keller, and L.~Gallmann, ``Water-window high harmonic generation with
  0.8-$\mu$m and 2.2-$\mu$m {OPCPA}s at 100 k{H}z,'' {\em Optics Express},
  vol.~29, no.~21, pp.~32996--33008, 2021.

\bibitem{gebhardt2021bright}
M.~Gebhardt, T.~Heuermann, R.~Klas, C.~Liu, A.~Kirsche, M.~Lenski, Z.~Wang,
  C.~Gaida, J.~Antonio-Lopez, A.~Sch{\"u}lzgen, {\em et~al.}, ``Bright,
  high-repetition-rate water window soft x-ray source enabled by nonlinear
  pulse self-compression in an antiresonant hollow-core fibre,'' {\em Light:
  Science \& Applications}, vol.~10, no.~1, pp.~1--7, 2021.

\bibitem{mero201843}
M.~Mero, Z.~Heiner, V.~Petrov, H.~Rottke, F.~Branchi, G.~M. Thomas, and M.~J.
  Vrakking, ``43 {W}, 1.55 $\mu$m and 12.5 {W}, 3.1 $\mu$m dual-beam, sub-10
  cycle, 100 k{H}z optical parametric chirped pulse amplifier,'' {\em Optics
  Letters}, vol.~43, no.~21, pp.~5246--5249, 2018.

\bibitem{nagy2019generation}
T.~Nagy, S.~H{\"a}drich, P.~Simon, A.~Blumenstein, N.~Walther, R.~Klas,
  J.~Buldt, H.~Stark, S.~Breitkopf, P.~J{\'o}j{\'a}rt, {\em et~al.},
  ``Generation of three-cycle multi-millijoule laser pulses at 318 {W} average
  power,'' {\em Optica}, vol.~6, no.~11, pp.~1423--1424, 2019.

\bibitem{storz2017parametric}
P.~Storz, J.~Tauch, M.~Wunram, A.~Leitenstorfer, and D.~Brida, ``Parametric
  amplification of phase-locked few-cycle pulses and ultraviolet harmonics
  generation in solids at high repetition rate,'' {\em Laser \& Photonics
  Reviews}, vol.~11, no.~6, p.~1700062, 2017.

\bibitem{zouflat}
X.~Zou, W.~Li, S.~Qu, K.~Liu, H.~Li, Q.~J. Wang, Y.~Zhang, and H.~Liang,
  ``Flat-top pumped multi-millijoule mid-infrared parametric chirped-pulse
  amplifier at 10 k{H}z repetition rate,'' {\em Laser \& Photonics Reviews},
  p.~2000292, 2021.

\bibitem{natile2019cep}
M.~Natile, A.~Golinelli, L.~Lavenu, F.~Guichard, M.~Hanna, Y.~Zaouter,
  R.~Chiche, X.~Chen, J.~Hergott, W.~Boutu, {\em et~al.}, ``{CEP}-stable
  high-energy ytterbium-doped fiber amplifier,'' {\em Optics Letters}, vol.~44,
  no.~16, pp.~3909--3912, 2019.

\bibitem{young2018roadmap}
L.~Young, K.~Ueda, M.~G{\"u}hr, P.~H. Bucksbaum, M.~Simon, S.~Mukamel,
  N.~Rohringer, K.~C. Prince, C.~Masciovecchio, M.~Meyer, {\em et~al.},
  ``Roadmap of ultrafast x-ray atomic and molecular physics,'' {\em The Journal
  of Physics B: Atomic, Molecular and Optical Physics}, vol.~51, no.~3,
  p.~032003, 2018.

\bibitem{toth2020sylos}
S.~Toth, T.~Stanislauskas, I.~Balciunas, R.~Budriunas, J.~Adamonis,
  R.~Danilevicius, K.~Viskontas, D.~Lengvinas, G.~Veitas, D.~Gadonas, {\em
  et~al.}, ``{SYLOS} lasers--the frontier of few-cycle, multi-{TW}, k{H}z
  lasers for ultrafast applications at extreme light infrastructure attosecond
  light pulse source,'' {\em Journal of Physics: Photonics}, vol.~2, no.~4,
  p.~045003, 2020.

\bibitem{mikaelsson2020high}
S.~Mikaelsson, J.~Vogelsang, C.~Guo, I.~Sytcevich, A.-L. Viotti, F.~Langer,
  Y.-C. Cheng, S.~Nandi, W.~Jin, A.~Olofsson, {\em et~al.}, ``A high-repetition
  rate attosecond light source for time-resolved coincidence spectroscopy,''
  {\em Nanophotonics}, vol.~1, no.~ahead-of-print, 2020.

\bibitem{goulielmakis2008single}
E.~Goulielmakis, M.~Schultze, M.~Hofstetter, V.~S. Yakovlev, J.~Gagnon,
  M.~Uiberacker, A.~L. Aquila, E.~Gullikson, D.~T. Attwood, R.~Kienberger, {\em
  et~al.}, ``Single-cycle nonlinear optics,'' {\em Science}, vol.~320,
  no.~5883, pp.~1614--1617, 2008.

\bibitem{takahashi2013attosecond}
E.~J. Takahashi, P.~Lan, O.~D. M{\"u}cke, Y.~Nabekawa, and K.~Midorikawa,
  ``Attosecond nonlinear optics using gigawatt-scale isolated attosecond
  pulses,'' {\em Nature Communications}, vol.~4, no.~1, pp.~1--9, 2013.

\bibitem{fabris2015synchronized}
D.~Fabris, T.~Witting, W.~Okell, D.~Walke, P.~Matia-Hernando, J.~Henkel,
  T.~Barillot, M.~Lein, J.~Marangos, and J.~Tisch, ``Synchronized pulses
  generated at 20 e{V} and 90 e{V} for attosecond pump--probe experiments,''
  {\em Nature Photonics}, vol.~9, no.~6, pp.~383--387, 2015.

\bibitem{manschwetus2016two}
B.~Manschwetus, L.~Rading, F.~Campi, S.~Maclot, H.~Coudert-Alteirac, J.~Lahl,
  H.~Wikmark, P.~Rudawski, C.~Heyl, B.~Farkas, {\em et~al.}, ``Two-photon
  double ionization of neon using an intense attosecond pulse train,'' {\em
  Physical Review A}, vol.~93, no.~6, p.~061402, 2016.

\bibitem{timmers2016polarization}
H.~Timmers, M.~Sabbar, J.~Hellwagner, Y.~Kobayashi, D.~M. Neumark, and S.~R.
  Leone, ``Polarization-assisted amplitude gating as a route to tunable,
  high-contrast attosecond pulses,'' {\em Optica}, vol.~3, no.~7, pp.~707--710,
  2016.

\bibitem{cousin2017attosecond}
S.~L. Cousin, N.~Di~Palo, B.~Buades, S.~M. Teichmann, M.~Reduzzi, M.~Devetta,
  A.~Kheifets, G.~Sansone, and J.~Biegert, ``Attosecond streaking in the water
  window: A new regime of attosecond pulse characterization,'' {\em Physical
  Review X}, vol.~7, no.~4, p.~041030, 2017.

\bibitem{li2019double}
J.~Li, A.~Chew, S.~Hu, J.~White, X.~Ren, S.~Han, Y.~Yin, Y.~Wang, Y.~Wu, and
  Z.~Chang, ``Double optical gating for generating high flux isolated
  attosecond pulses in the soft x-ray regime,'' {\em Optics Express}, vol.~27,
  no.~21, pp.~30280--30286, 2019.

\bibitem{ye2020attosecond}
P.~Ye, T.~Csizmadia, L.~G. Oldal, H.~N. Gopalakrishna, M.~F{\"u}le, Z.~Filus,
  B.~Nagyill{\'e}s, Z.~Div{\'e}ki, T.~Gr{\'o}sz, M.~Dumergue, {\em et~al.},
  ``Attosecond pulse generation at {ELI-ALPS} 100 {kHz} repetition rate
  beamline,'' {\em Journal of Physics B: Atomic, Molecular and Optical
  Physics}, vol.~53, no.~15, p.~154004, 2020.

\bibitem{osolodkov2020generation}
M.~Osolodkov, F.~J. Furch, F.~Schell, P.~{\v{S}}u{\v{s}}njar, F.~Cavalcante,
  C.~S. Menoni, C.~P. Schulz, T.~Witting, and M.~J. Vrakking, ``Generation and
  characterisation of few-pulse attosecond pulse trains at 100 {kHz} repetition
  rate,'' {\em Journal of Physics B: Atomic, Molecular and Optical Physics},
  vol.~53, no.~19, p.~194003, 2020.

\bibitem{makos2020alpha}
I.~Makos, I.~Orfanos, A.~Nayak, J.~Peschel, B.~Major, I.~Liontos,
  E.~Skantzakis, N.~Papadakis, C.~Kalpouzos, M.~Dumergue, {\em et~al.}, ``A
  10-gigawatt attosecond source for non-linear {XUV} optics and
  {XUV}-pump-{XUV}-probe studies,'' {\em Scientific Reports}, vol.~10, no.~1,
  pp.~1--18, 2020.

\bibitem{witting2021generation}
T.~Witting, M.~Osolodkov, F.~Schell, F.~Morales, S.~Patchkovskii, P.~Susnjar,
  F.~Cavalcante, C.~S. Menoni, C.~P. Schulz, F.~J. Furch, {\em et~al.},
  ``Generation and characterisation of isolated attosecond pulses at 100 k{H}z
  repetition rate,'' {\em Optica, accepted, DOI 10.1364/OPTICA.443521}, 2021.

\bibitem{cingoz2012direct}
A.~Cing{\"o}z, D.~C. Yost, T.~K. Allison, A.~Ruehl, M.~E. Fermann, I.~Hartl,
  and J.~Ye, ``Direct frequency comb spectroscopy in the extreme ultraviolet,''
  {\em Nature}, vol.~482, no.~7383, pp.~68--71, 2012.

\bibitem{pupeza2013compact}
I.~Pupeza, S.~Holzberger, T.~Eidam, H.~Carstens, D.~Esser, J.~Weitenberg,
  P.~Ru{\ss}b{\"u}ldt, J.~Rauschenberger, J.~Limpert, T.~Udem, {\em et~al.},
  ``Compact high-repetition-rate source of coherent 100 e{V} radiation,'' {\em
  Nature Photonics}, vol.~7, no.~8, pp.~608--612, 2013.

\bibitem{ozawa2015high}
A.~Ozawa, Z.~Zhao, M.~Kuwata-Gonokami, and Y.~Kobayashi, ``High average power
  coherent vuv generation at 10 {MHz} repetition frequency by intracavity high
  harmonic generation,'' {\em Optics Express}, vol.~23, no.~12,
  pp.~15107--15118, 2015.

\bibitem{porat2018phase}
G.~Porat, C.~M. Heyl, S.~B. Schoun, C.~Benko, N.~D{\"o}rre, K.~L. Corwin, and
  J.~Ye, ``Phase-matched extreme-ultraviolet frequency-comb generation,'' {\em
  Nature Photonics}, vol.~12, no.~7, pp.~387--391, 2018.

\bibitem{saule2019high}
T.~Saule, S.~Heinrich, J.~Sch{\"o}tz, N.~Lilienfein, M.~H{\"o}gner, O.~deVries,
  M.~Pl{\"o}tner, J.~Weitenberg, D.~Esser, J.~Schulte, {\em et~al.},
  ``High-flux ultrafast extreme-ultraviolet photoemission spectroscopy at 18.4
  {MHz} pulse repetition rate,'' {\em Nature Communications}, vol.~10, no.~1,
  pp.~1--10, 2019.

\bibitem{lee2011optimizing}
J.~Lee, D.~R. Carlson, and R.~J. Jones, ``Optimizing intracavity high harmonic
  generation for xuv fs frequency combs,'' {\em Optics Express}, vol.~19,
  no.~23, pp.~23315--23326, 2011.

\bibitem{carstens2016high}
H.~Carstens, M.~H{\"o}gner, T.~Saule, S.~Holzberger, N.~Lilienfein,
  A.~Guggenmos, C.~Jocher, T.~Eidam, D.~Esser, V.~Tosa, {\em et~al.},
  ``High-harmonic generation at 250 {MHz} with photon energies exceeding 100
  e{V},'' {\em Optica}, vol.~3, no.~4, pp.~366--369, 2016.

\bibitem{corder2018ultrafast}
C.~Corder, P.~Zhao, J.~Bakalis, X.~Li, M.~D. Kershis, A.~R. Muraca, M.~G.
  White, and T.~K. Allison, ``Ultrafast extreme ultraviolet photoemission
  without space charge,'' {\em Structural Dynamics}, vol.~5, no.~5, p.~054301,
  2018.

\bibitem{FilusZoltan2022}
Z.~Filus, ``Liquid-cooled modular gas target cell system for high-order
  harmonic generation using high average power laser systems,'' {\em In
  preparation}, 2021.

\bibitem{siegman1986lasers}
A.~E. Siegman, ``Lasers,'' {\em Mill Valley, CA}, vol.~37, no.~208, p.~661,
  1986.

\bibitem{ye2014minimizing}
P.~Ye, H.~Teng, X.-K. He, S.-Y. Zhong, L.-F. Wang, M.-J. Zhan, W.~Zhang, C.-X.
  Yun, and Z.-Y. Wei, ``Minimizing the angular divergence of high-order
  harmonics by truncating the truncated bessel beam,'' {\em Physical Review A},
  vol.~90, no.~6, p.~063808, 2014.

\bibitem{peatross1994high}
J.~Peatross, J.~Chaloupka, and D.~Meyerhofer, ``High-order harmonic generation
  with an annular laser beam,'' {\em Optics Letters}, vol.~19, no.~13,
  pp.~942--944, 1994.

\bibitem{kuhn2017eli}
S.~K{\"u}hn, M.~Dumergue, S.~Kahaly, S.~Mondal, M.~F{\"u}le, T.~Csizmadia,
  B.~Farkas, B.~Major, Z.~V{\'a}rallyay, E.~Cormier, {\em et~al.}, ``The
  {ELI-ALPS} facility: the next generation of attosecond sources,'' {\em The
  Journal of Physics B: Atomic, Molecular and Optical Physics}, vol.~50,
  no.~13, p.~132002, 2017.

\bibitem{mairesse2003attosecond}
Y.~Mairesse, A.~De~Bohan, L.~Frasinski, H.~Merdji, L.~Dinu, P.~Monchicourt,
  P.~Breger, M.~Kova{\v{c}}ev, R.~Ta{\"\i}eb, B.~Carr{\'e}, {\em et~al.},
  ``Attosecond synchronization of high-harmonic soft x-rays,'' {\em Science},
  vol.~302, no.~5650, pp.~1540--1543, 2003.

\bibitem{gaumnitz2018extreme}
T.~Gaumnitz, A.~Jain, and H.~J. W{\"o}rner, ``Extreme-ultraviolet high-order
  harmonic generation from few-cycle annular beams,'' {\em Optics Letters},
  vol.~43, no.~18, pp.~4506--4509, 2018.

\bibitem{chang2016fundamentals}
Z.~Chang, {\em Fundamentals of attosecond optics}.
\newblock CRC press, 2016.

\bibitem{rivas2018propagation}
D.~Rivas, B.~Major, M.~Weidman, W.~Helml, G.~Marcus, R.~Kienberger,
  D.~Charalambidis, P.~Tzallas, E.~Balogh, K.~Kov{\'a}cs, {\em et~al.},
  ``Propagation-enhanced generation of intense high-harmonic continua in the
  100-e{V} spectral region,'' {\em Optica}, vol.~5, no.~10, pp.~1283--1289,
  2018.

\bibitem{major2019effect}
B.~Major, K.~Kov{\'a}cs, V.~Tosa, P.~Rudawski, A.~L’Huillier, and
  K.~Varj{\'u}, ``Effect of plasma-core-induced self-guiding on phase matching
  of high-order harmonic generation in gases,'' {\em Journal of the Optical
  Society of America B}, vol.~36, no.~6, pp.~1594--1601, 2019.

\bibitem{major2020propagation}
B.~Major, M.~Kretschmar, O.~Ghafur, A.~Hoffmann, K.~Kov{\'a}cs, K.~Varj{\'u},
  B.~Senfftleben, J.~T{\"u}mmler, I.~Will, T.~Nagy, {\em et~al.},
  ``Propagation-assisted generation of intense few-femtosecond high-harmonic
  pulses,'' {\em Journal of Physics: Photonics}, vol.~2, no.~3, p.~034002,
  2020.

\bibitem{jin2020optimal}
C.~Jin, X.~Tang, B.~Li, K.~Wang, and C.~Lin, ``Optimal spatial separation of
  high-order harmonics from infrared driving lasers with an annular beam in the
  overdriven regime,'' {\em Physical Review Applied}, vol.~14, no.~1,
  p.~014057, 2020.

\bibitem{shestaev2020high}
E.~Shestaev, D.~Hoff, A.~Sayler, A.~Klenke, S.~H{\"a}drich, F.~Just, T.~Eidam,
  P.~J{\'o}j{\'a}rt, Z.~V{\'a}rallyay, K.~Osvay, {\em et~al.}, ``High-power
  ytterbium-doped fiber laser delivering few-cycle, carrier-envelope
  phase-stable 100 $\mu${J} pulses at 100 k{H}z,'' {\em Optics Letters},
  vol.~45, no.~1, pp.~97--100, 2020.

\end{thebibliography}
	
\end{document}